\pdfoutput=1

\documentclass[a4paper,12pt]{elsarticle}
\usepackage{slashed}
\usepackage{amssymb,amsmath,amsbsy}
\usepackage{mathrsfs}
\usepackage{bm}
\usepackage{hyperref}
\usepackage{mdframed}
\usepackage[top=1.2in,bottom=1.2in,left=0.9in,includefoot]{geometry}
\usepackage{xcolor}

\newcounter{bla}

\def\sfr{{\tt SmeftFR}~}
\def\frules{{\tt FeynRules}~}
\def\webpage{\url{www.fuw.edu.pl/smeft}}

\def\eq#1{eq.~(\ref{#1})}

\def\Ref#1{ref.~\cite{#1}}

\newcommand{\bea}{\begin{eqnarray}}
\newcommand{\eea}{\end{eqnarray}}
\newcommand{\nn}{\nonumber\\}

\newcommand{\bce}{\begin{center}}
\newcommand{\ece}{\end{center}}
\newcommand{\be}{\begin{equation}}
\newcommand{\ee }{\end{equation}}
\newcommand{\btb}{\begin{tabular}}
\newcommand{\etb}{\end{tabular}}

\newcommand{\eps}{\varepsilon}
\newcommand{\vp}{\varphi}

\def\lcal{{\cal L}}

\def\wt{\widetilde}

\newcommand{\tvp}{\widetilde{\varphi}}
\newcommand{\vpj }{\mbox{${\vp^\dag i\,\raisebox{2mm}{\boldmath
        ${}^\leftrightarrow$}\hspace{-4mm} D_\mu\,\vp}$}}
\newcommand{\vpjt}{\mbox{${\vp^\dag i\,\raisebox{2mm}{\boldmath
        ${}^\leftrightarrow$}\hspace{-4mm} D_\mu^{\,I}\,\vp}$}}

\usepackage{color}

\definecolor{orange}{rgb}{0.9,0.2,0}
\definecolor{brown}{rgb}{0.7,0.3,0.2}
\definecolor{fuxia}{rgb}{1,0,1}
\definecolor{skyblue}{rgb}{0,0.1,0.9}
\definecolor{violetred}{rgb}{0.8,0.13,0.56}
\definecolor{deeppink}{rgb}{1.00,0.08,0.5}
\definecolor{pink}{rgb}{1.00,0.75,0.80}
\definecolor{orchid}{rgb}{0.85,0.44,0.84}
\definecolor{lightpink}{rgb}{1.00,0.71,0.76}
\definecolor{bluish}{rgb}{0,0.6,0.8}
\newcommand{\red}{\textcolor{red}}

\numberwithin{equation}{section}

\journal{Computer Physics Communications}

\begin{document}

\begin{frontmatter}

\title{\sfr -- Feynman rules generator for the Standard Model Effective
  Field Theory}

\author[a]{A. Dedes}
\author[b]{M. Paraskevas}
\author[b]{J. Rosiek\corref{author}}
\author[a]{K. Suxho}
\author[a]{L. Trifyllis}

\cortext[author]{Corresponding author, e-mail address
    \href{mailto:janusz.rosiek@fuw.edu.pl}{\tt janusz.rosiek@fuw.edu.pl}}

\address[a]{Department of Physics, University of Ioannina, GR 45110,
  Ioannina, Greece}

\address[b]{Faculty of Physics, University of Warsaw, Pasteura 5, 02-093
  Warsaw, Poland}

\begin{abstract}
  We present {\tt SmeftFR}, a Mathematica package designed to generate
  the Feynman rules for the Standard Model Effective Field Theory
  (SMEFT) including the complete set of gauge invariant operators up
  to dimension~6.  Feynman rules are generated with the use of \frules
  package, directly in the physical (mass eigenstates) basis for all
  fields.  The complete set of interaction vertices can be derived
  including all or any chosen subset of SMEFT operators. As an option,
  the user can also choose preferred gauge fixing, generating Feynman
  rules in unitary or $R_\xi$-gauges (the latter include generation of
  ghost vertices).  Further options allow to treat neutrino fields as
  massless Weyl or massive Majorana fermions. The derived Lagrangian
  in the mass basis can be exported in various formats supported by
  {\tt FeynRules}, such as {\tt UFO}, {\tt FeynArts}, \textit{etc}.
  Initialisation of numerical values of $d=6$ Wilson coefficients used
  by \sfr is interfaced to WCxf format.  The package also includes
  dedicated Latex generator allowing to print the result in clear
  human-readable form.  \sfr can be downloaded from the
  address \webpage.
\end{abstract}

\begin{keyword}
Standard Model Effective Field Theory\sep Feynman rules \sep unitary
and $R_\xi$-gauges
\end{keyword}

\end{frontmatter}

\newpage

\noindent {\bf PROGRAM SUMMARY}

\begin{small}
\noindent
{\em Manuscript Title:} \\ 
\sfr -- Feynman rules generator for the Standard Model Effective Field
Theory\\
{\em Authors:} Athanasios Dedes, Michael Paraskevas, Janusz Rosiek,
Kristaq Suxho, Lampros Trifyllis\\
{\em Program Title:} \sfr v2.0 \\
{\em Journal Reference:} \\
{\em Catalogue identifier:} \\
{\em Licensing provisions:} None \\
{\em Programming language:} Mathematica 11.3 (earlier versions were
reported to have problems running this code) \\
{\em Computer:} any running Mathematica \\
{\em Operating system:} any running Mathematica \\
{\em RAM:} allocated dynamically by Mathematica, at least 4GB total
RAM suggested \\ 
{\em Number of processors used:} allocated dynamically by Mathematica \\
{\em Supplementary material:} None \\
{\em Keywords:} Standard Model Effective Field Theory, Feynman rules,
unitary and $R_\xi$ gauges \\
{\em Classification:}\\ \begin{tabular}{ll}
11.1 & General, High Energy Physics and Computing,\\
4.4 & Feynman diagrams,\\
5 & Computer Algebra.  \\
\end{tabular}\\
{\em External routines/libraries:} Wolfram Mathematica program \\
{\em Subprograms used:} {\tt FeynRules v2.3} package \\
{\em Nature of problem:}\\ 
Automatised generation of Feynman rules in physical (mass) basis for
the Standard Model Effective Field Theory with user defined operator
subset and gauge fixing terms selection.  \\
{\em Solution method:}\\ 
Implementation (as the Mathematica package) of the results of Ref.~[1]
in the \frules package~[2], including dynamic ``model files''
generation.  \\
{\em Restrictions:} None\\
{\em Unusual features:} None \\
{\em Additional comments:} None \\
{\em Running time:} depending on control variable settings, from few
minutes for a selected subset of few SMEFT operators and Feynman rules
generation up to several hours for generating UFO output for maximum
60-operator set (using Mathematica 11.3 running on a personal
computer) \\

\end{small}

\newpage

\section{Introduction}
\label{sec:intro}
  
Despite lack of direct discoveries of new particles at LHC, indirect
experiments searching for cosmological dark matter and dark energy,
neutrino masses and mixing, heavy meson decays, anomalous magnetic
moment of muon, point to existence of new physics phenomena that go
beyond the Standard Model (SM)~\cite{Weinberg:1967tq, Glashow, Salam}
of particle physics.  It has become customary in recent years to
parameterise the potential effects of such New Physics (NP) in terms
of the so-called SM Effective Field Theory
(SMEFT)~\cite{Weinberg:1980wa, Coleman:1969sm,
  Callan:1969sn}\footnote{For a recent review see
  ref.~\cite{Manohar:2018aog}.}.  In this theory, SM is extended by a
complete set of gauge invariant operators constructed only by the SM
fields which are considered to be light in mass with respect to, yet
unknown, particles responsible for NP phenomena.  Such
(non-renormalisable) operators can be classified according to their
growing mass dimension, with couplings (usually called Wilson
coefficients) suppressed by respectively inverse powers of a typical
mass scale $\Lambda$ of the NP extension.

As observable effects of new operators are typically suppressed by
powers of $v/\Lambda$ (where $v$ is the true vacuum expectation value
of the Higgs field), in many analyses it is sufficient to restrict the
maximal dimension of new operators to $d\le 6$.  Classification of the
complete independent set of the gauge invariant operators up to $d=6$
was conducted in an older study by Buchm\"uller and
Wyler~\cite{Buchmuller:1985jz} and more recently put in a
non-redundant form in ref.~\cite{Grzadkowski:2010es}, where the basis
of such operators, usually referred as the ``Warsaw basis'', has been
given.  Suppressing the flavour indices of the fields and not counting
hermitian conjugated operators, Warsaw basis contains 59+1
baryon-number conserving and 4 baryon-number violating operators.

In its most general version, SMEFT is a hugely complicated model.
Including all possible CP- and flavour-violating interactions, it
contains 2499 free parameters.  Due to large number and complicated
structure of new terms in the Lagrangian, theoretical calculations of
physical processes within the SMEFT can be very challenging --- it is
enough to notice that the number of primary vertices when SMEFT is
quantised in $R_\xi$-gauges, printed for the first time in
ref.~\cite{Dedes:2017zog}, is almost 400 without counting the
hermitian conjugates.  Thus, it is important to develop technical
methods and tools facilitating such calculations, starting from
developing the universal set of the Feynman rules for propagators and
vertices for physical fields, after spontaneous symmetry breaking
(SSB) of the full effective theory. The initial version of relevant
package, \sfr v1, was announced and briefly described for the first
time in Appendix B of ref.~\cite{Dedes:2017zog}.  In this paper we
present \sfr v2.0, a {\em Mathematica} symbolic language package
generating Feynman rules in several formats, based on the formulae
developed in ref.~\cite{Dedes:2017zog}.  Comparing to version 1, the
new code has been extended with a number of additional options and
capabilities. It has also been made and tested to be compatible with
many other publicly available high-energy physics related computer
codes accepting standardised input and output data formats.  Features
of the presented package contain:
\begin{itemize}
\item \sfr is able to generate interactions in the most general form
  of the SMEFT Lagrangian, without any restrictions on the structure
  of flavour violating terms and on CP-, lepton- or baryon-number
  conservation.\footnote{However, we do restrict ourselves to linear
    realisations of the SSB.}  Feynman rules are expressed in terms of
  physical SM fields and canonically normalised Goldstone and ghost
  fields.  Expressions for interaction vertices are analytically
  expanded in powers of inverse New Physics scale $1/\Lambda$, with
  all terms of dimension higher than $d=6$ consistently truncated.
\item \sfr is written as an overlay to \frules
  package~\cite{Alloul:2013bka}, used as the engine to generate
  Feynman rules.
\item Including the full set of SMEFT parameters in model files for
  \frules may lead to very slow computations.  \sfr can generate
  \frules model files dynamically, including only the user defined
  subset of higher dimension operators.  It significantly speeds up
  the calculations and produces simpler final result, containing only
  the Wilson coefficients relevant for a process chosen to analyse.
\item Feynman rules can be generated in the unitary or in linear
  $R_\xi$-gauges by exploiting four different gauge-fixing parameters
  $\xi_\gamma, \xi_Z, \xi_W, \xi_G$ for thorough amplitude checks.  In
  the latter case also all relevant ghost vertices are obtained.
\item Feynman rules are calculated first in {\em Mathematica}/\frules
  format.  They can be further exported in other formats: {\tt
    UFO}~\cite{Degrande:2011ua} (importable to Monte Carlo generators
  like {\tt MadGraph5\_aMC@NLO 5} ~\cite{Alwall:2014hca}, {\tt
    Sherpa}~\cite{Gleisberg:2008ta}, {\tt
    CalcHEP}~\cite{Belyaev:2012qa}, {\tt Whizard}~\cite{Kilian:2007gr,
    Christensen:2010wz}), {\tt FeynArts}~\cite{Hahn:2000kx} which
  generates inputs for loop amplitude calculators like {\tt
    FeynCalc}~\cite{Shtabovenko:2016sxi}, or {\tt
    FormCalc}~\cite{Hahn:2010zi}, and others output types supported by
  \frules.
\item \sfr provides a dedicated Latex~generator, allowing to display
  vertices and analytical expressions for Feynman rules in clear human
  readable form, best suited for hand-made calculations.
\item \sfr is interfaced to the WCxf format~\cite{Aebischer:2017ugx}
  of Wilson coefficients. Numerical values of SMEFT parameters in
  model files can be read from WCxf JSON-type input produced by other
  computer packages written for SMEFT.  Alternatively, \sfr can
  translate \frules model files to the WCxf format.
\item Further package options allow to treat neutrino fields as
  massless Weyl or (in the case of non-vanishing dimension-5 operator)
  massive Majorana fermions, to correct signs in 4-fermion
  interactions not yet fully supported by \frules and to perform some
  additional operations as described later in this manual.
\end{itemize}
Feynman rules derived in ref.~\cite{Dedes:2017zog} using the {\tt
  SmeftFR} package have been used successfully in many articles
including refs.~\cite{Dedes:2018seb, Dedes:2019bew, Dawson:2018pyl,
  Vryonidou:2018eyv, Hesari:2018ssq, Dawson:2018liq, Dawson:2018jlg,
  Baglio:2018bkm, Dawson:2018dxp, Silvestrini:2018dos,
  Neumann:2019kvk} and have passed certain non-trivial tests, such as
gauge-fixing parameter independence of the $S$-matrix elements,
validity of Ward identities, cancellation of infinities in loop
calculations, \textit{etc}.

We note here in passing, that there is a growing number of publicly
available codes performing computations related to SMEFT.  These
include, {\tt Wilson}~\cite{Aebischer:2018bkb}, {\tt
  DSixTools}~\cite{Celis:2017hod}, {\tt
  MatchingTools}~\cite{Criado:2017khh}, which are codes for running
and matching Wilson coefficients, {\tt
  SMEFTsim}~\cite{Brivio:2017btx}, a package for calculating tree
level observables, {\tt CoDEx}~\cite{Bakshi:2018ics} or a version of
{\tt SARAH} code~\cite{Gabelmann:2018axh}, that calculate Wilson
Coefficients after the decoupling of a more fundamental theory, and
finally, {\tt DirectDM}~\cite{Bishara:2017nnn}, a code for dark matter
EFT.  To a degree, these codes (especially the ones supporting WCxf
format) can be used in conjunction with {\tt SmeftFR}.  For example,
some of them can provide the numerical input for Wilson coefficients
of higher dimensional operators at scale $\Lambda$, while others, the
running of these coefficients from that scale down to the EW one.
Alternatively, Feynman rules evaluated by \sfr can be used with Monte
Carlo generators to test the predictions of other packages.

The paper is organised as follows.  After this Introduction, in
Sec.~\ref{sec:basis} we define the notation and conventions, listing
for reference the operator set in Warsaw basis and the formulae for
transition to the mass basis.  In Sec.~\ref{sec:sfr}, we present the
structure of the code, installation procedure and available functions.
Section~\ref{sec:sample}, contains examples of programs generating the
Feynman rules in various formats.  We conclude in
Sec.~\ref{sec:summary}.

\section{SMEFT Lagrangian in Warsaw and mass basis}
\label{sec:basis}

The classification of higher order operators in SMEFT is done in terms
of fields in electroweak basis, before the Spontaneous Symmetry
Breaking.  \sfr uses the so-called ``Warsaw
basis''~\cite{Grzadkowski:2010es} as a starting point to calculate
physical states in SMEFT and their interactions.  For easier reference
we copy here from ref.~\cite{Grzadkowski:2010es}
Tables~\ref{tab:no4ferm} and~\ref{tab:4ferm} containing the full
collection of $d=6$ operators in Warsaw basis.\footnote{We do not list
  here all details of conventions used --- they are identical to these
  listed in refs.~\cite{Grzadkowski:2010es, Dedes:2017zog}.} In
addition, we include the single lepton flavour violating dimension-5
operator:
\be
\label{qnunu}
Q_{\nu\nu} = \eps_{jk} \eps_{mn} \vp^j \vp^m (l^{\prime k}_{Lp})^T \,
\mathbb{C}\, l^{\prime n}_{Lr} ~\equiv~ (\tvp^\dag l'_{Lp})^T\, \mathbb{C}\,
(\tvp^\dag l'_{Lr})\;,
\ee
where $\mathbb{C}$ is the charge conjugation matrix.

\begin{table}[t] 
\centering \renewcommand{\arraystretch}{1.5} \btb{||c|c||c|c||c|c||}
\hline \hline \multicolumn{2}{||c||}{$X^3$} &
\multicolumn{2}{|c||}{$\vp^6$~ and~ $\vp^4 D^2$} &
\multicolumn{2}{|c||}{$\psi^2\vp^3$}\\ \hline
$Q_G$ & $f^{ABC} G_\mu^{A\nu} G_\nu^{B\rho} G_\rho^{C\mu} $ & $Q_\vp$
& $(\vp^\dag\vp)^3$ & $Q_{e\vp}$ & $(\vp^\dag \vp)(\bar l'_p e'_r
\vp)$\\
$Q_{\wt G}$ & $f^{ABC} \wt G_\mu^{A\nu} G_\nu^{B\rho} G_\rho^{C\mu} $
& $Q_{\vp\Box}$ & $(\vp^\dag \vp)\raisebox{-.5mm}{$\Box$}(\vp^\dag
\vp)$ & $Q_{u\vp}$ & $(\vp^\dag \vp)(\bar q'_p u'_r \tvp)$\\
$Q_W$ & $\eps^{IJK} W_\mu^{I\nu} W_\nu^{J\rho} W_\rho^{K\mu}$ &
$Q_{\vp D}$ & $\left(\vp^\dag D^\mu\vp\right)^* \left(\vp^\dag
D_\mu\vp\right)$ & $Q_{d\vp}$ & $(\vp^\dag \vp)(\bar q'_p d'_r \vp)$\\
$Q_{\wt W}$ & $\eps^{IJK} \wt W_\mu^{I\nu} W_\nu^{J\rho}
W_\rho^{K\mu}$ &&&&\\
\hline \hline \multicolumn{2}{||c||}{$X^2\vp^2$} &
\multicolumn{2}{|c||}{$\psi^2 X\vp$} &
\multicolumn{2}{|c||}{$\psi^2\vp^2 D$}\\ \hline $Q_{\vp G}$ &
$\vp^\dag \vp\, G^A_{\mu\nu} G^{A\mu\nu}$ & $Q_{eW}$ & $(\bar l'_p
\sigma^{\mu\nu} e'_r) \tau^I \vp W_{\mu\nu}^I$ & $Q_{\vp l}^{(1)}$ &
$(\vpj)(\bar l'_p \gamma^\mu l'_r)$\\
$Q_{\vp\wt G}$ & $\vp^\dag \vp\, \wt G^A_{\mu\nu} G^{A\mu\nu}$ &
$Q_{eB}$ & $(\bar l'_p \sigma^{\mu\nu} e'_r) \vp B_{\mu\nu}$ & $Q_{\vp
  l}^{(3)}$ & $(\vpjt)(\bar l'_p \tau^I \gamma^\mu l'_r)$\\
$Q_{\vp W}$ & $\vp^\dag \vp\, W^I_{\mu\nu} W^{I\mu\nu}$ & $Q_{uG}$ &
$(\bar q^\prime_p \sigma^{\mu\nu} \mathcal{T}^A u^\prime_r) \tvp\,
G_{\mu\nu}^A$ & $Q_{\vp e}$ & $(\vpj)(\bar e'_p \gamma^\mu e'_r)$\\
$Q_{\vp\wt W}$ & $\vp^\dag \vp\, \wt W^I_{\mu\nu} W^{I\mu\nu}$ &
$Q_{uW}$ & $(\bar q'_p \sigma^{\mu\nu} u'_r) \tau^I \tvp\,
W_{\mu\nu}^I$ & $Q_{\vp q}^{(1)}$ & $(\vpj)(\bar q'_p \gamma^\mu
q'_r)$\\
$Q_{\vp B}$ & $ \vp^\dag \vp\, B_{\mu\nu} B^{\mu\nu}$ & $Q_{uB}$ &
$(\bar q'_p \sigma^{\mu\nu} u'_r) \tvp\, B_{\mu\nu}$& $Q_{\vp
  q}^{(3)}$ & $(\vpjt)(\bar q'_p \tau^I \gamma^\mu q'_r)$\\
$Q_{\vp\wt B}$ & $\vp^\dag \vp\, \wt B_{\mu\nu} B^{\mu\nu}$ & $Q_{dG}$
& $(\bar q'_p \sigma^{\mu\nu} \mathcal{T}^A d'_r) \vp\, G_{\mu\nu}^A$
& $Q_{\vp u}$ & $(\vpj)(\bar u'_p \gamma^\mu u'_r)$\\
$Q_{\vp WB}$ & $ \vp^\dag \tau^I \vp\, W^I_{\mu\nu} B^{\mu\nu}$ &
$Q_{dW}$ & $(\bar q'_p \sigma^{\mu\nu} d'_r) \tau^I \vp\,
W_{\mu\nu}^I$ & $Q_{\vp d}$ & $(\vpj)(\bar d'_p \gamma^\mu d'_r)$\\
$Q_{\vp\wt WB}$ & $\vp^\dag \tau^I \vp\, \wt W^I_{\mu\nu} B^{\mu\nu}$
& $Q_{dB}$ & $(\bar q'_p \sigma^{\mu\nu} d'_r) \vp\, B_{\mu\nu}$ &
$Q_{\vp u d}$ & $i(\tvp^\dag D_\mu \vp)(\bar u'_p \gamma^\mu
d'_r)$\\ \hline \hline \etb
\caption{ Dimension-6 operators other than the four-fermion ones (from
  \Ref{Grzadkowski:2010es}).  For brevity we suppress fermion chiral
  indices $L,R$.}
   \label{tab:no4ferm}
\end{table}

\begin{table}[htb!]
\centering \renewcommand{\arraystretch}{1.5}
\begin{tabular}{||c|c||c|c||c|c||}
\hline\hline \multicolumn{2}{||c||}{$(\bar LL)(\bar LL)$} &
\multicolumn{2}{|c||}{$(\bar RR)(\bar RR)$} &
\multicolumn{2}{|c||}{$(\bar LL)(\bar RR)$}\\ \hline $Q_{ll}$ & $(\bar
l'_p \gamma_\mu l'_r)(\bar l'_s \gamma^\mu l'_t)$ & $Q_{ee}$ & $(\bar
e'_p \gamma_\mu e'_r)(\bar e'_s \gamma^\mu e'_t)$ & $Q_{le}$ & $(\bar
l'_p \gamma_\mu l'_r)(\bar e'_s \gamma^\mu e'_t)$ \\ $Q_{qq}^{(1)}$ &
$(\bar q'_p \gamma_\mu q'_r)(\bar q'_s \gamma^\mu q'_t)$ & $Q_{uu}$ &
$(\bar u'_p \gamma_\mu u'_r)(\bar u'_s \gamma^\mu u'_t)$ & $Q_{lu}$ &
$(\bar l'_p \gamma_\mu l'_r)(\bar u'_s \gamma^\mu u'_t)$
\\ $Q_{qq}^{(3)}$ & $(\bar q'_p \gamma_\mu \tau^I q'_r)(\bar q'_s
\gamma^\mu \tau^I q'_t)$ & $Q_{dd}$ & $(\bar d'_p \gamma_\mu
d'_r)(\bar d'_s \gamma^\mu d'_t)$ & $Q_{ld}$ & $(\bar l'_p \gamma_\mu
l'_r)(\bar d'_s \gamma^\mu d'_t)$ \\ $Q_{lq}^{(1)}$ & $(\bar l'_p
\gamma_\mu l'_r)(\bar q'_s \gamma^\mu q'_t)$ & $Q_{eu}$ & $(\bar e'_p
\gamma_\mu e'_r)(\bar u'_s \gamma^\mu u'_t)$ & $Q_{qe}$ & $(\bar q'_p
\gamma_\mu q'_r)(\bar e'_s \gamma^\mu e'_t)$ \\ $Q_{lq}^{(3)}$ &
$(\bar l'_p \gamma_\mu \tau^I l'_r)(\bar q'_s \gamma^\mu \tau^I q'_t)$
& $Q_{ed}$ & $(\bar e'_p \gamma_\mu e'_r)(\bar d'_s\gamma^\mu d'_t)$ &
$Q_{qu}^{(1)}$ & $(\bar q'_p \gamma_\mu q'_r)(\bar u'_s \gamma^\mu
u'_t)$ \\ && $Q_{ud}^{(1)}$ & $(\bar u'_p \gamma_\mu u'_r)(\bar d'_s
\gamma^\mu d'_t)$ & $Q_{qu}^{(8)}$ & $(\bar q'_p \gamma_\mu
\mathcal{T}^A q'_r)(\bar u'_s \gamma^\mu \mathcal{T}^A u'_t)$ \\ &&
$Q_{ud}^{(8)}$ & $(\bar u'_p \gamma_\mu \mathcal{T}^A u'_r)(\bar d'_s
\gamma^\mu \mathcal{T}^A d'_t)$ & $Q_{qd}^{(1)}$ & $(\bar q'_p
\gamma_\mu q'_r)(\bar d'_s \gamma^\mu d'_t)$ \\ &&&& $Q_{qd}^{(8)}$ &
$(\bar q'_p \gamma_\mu \mathcal{T}^A q'_r)(\bar d'_s \gamma^\mu
\mathcal{T}^A d'_t)$\\ \hline\hline \multicolumn{2}{||c||}{$(\bar
  LR)(\bar RL)$ and $(\bar LR)(\bar LR)$} &
\multicolumn{4}{|c||}{$B$-violating}\\\hline $Q_{ledq}$ & $(\bar
l_p^{'j}e'_r)(\bar d'_s q_t^{'j})$ & $Q_{duq}$ &
\multicolumn{3}{|c||}{$\eps^{\alpha\beta\gamma} \eps_{jk} \left[
    (d^{'\alpha}_p)^T \mathbb{C} u^{'\beta}_r \right]\left[(q^{'\gamma
      j}_s)^T \mathbb{C} l^{'k}_t\right]$}\\ $Q_{quqd}^{(1)}$ & $(\bar
q_p^{'j} u'_r) \eps_{jk} (\bar q_s^{'k} d'_t)$ & $Q_{qqu}$ &
\multicolumn{3}{|c||}{$\eps^{\alpha\beta\gamma} \eps_{jk} \left[
    (q^{'\alpha j}_p)^T \mathbb{C} q^{'\beta k}_r
    \right]\left[(u^{'\gamma}_s)^T \mathbb{C}
    e'_t\right]$}\\ $Q_{quqd}^{(8)}$ & $(\bar q_p^{'j} \mathcal{T}^A
u'_r) \eps_{jk} (\bar q_s^{'k} \mathcal{T}^A d'_t)$ & $Q_{qqq}$ &
\multicolumn{3}{|c||}{$\eps^{\alpha\beta\gamma} \eps_{jn} \eps_{km}
  \left[ (q^{'\alpha j}_p)^T \mathbb{C} q^{'\beta k}_r
    \right]\left[(q^{'\gamma m}_s)^T \mathbb{C}
    l^{'n}_t\right]$}\\ $Q_{lequ}^{(1)}$ & $(\bar l_p^{\prime\, j}
e^\prime_r) \eps_{jk} (\bar q_s^{\prime\, k} u^\prime_t)$ & $Q_{duu}$
& \multicolumn{3}{|c||}{$\eps^{\alpha\beta\gamma} \left[
    (d^{'\alpha}_p)^T \mathbb{C} u^{'\beta}_r
    \right]\left[(u^{'\gamma}_s)^T \mathbb{C}
    e'_t\right]$}\\ $Q_{lequ}^{(3)}$ & $(\bar l_p^{'j} \sigma_{\mu\nu}
e'_r) \eps_{jk} (\bar q_s^{'k} \sigma^{\mu\nu} u'_t)$ & &
\multicolumn{3}{|c||}{}\\ \hline\hline
\end{tabular}
\caption{Four-fermion operators (from \Ref{Grzadkowski:2010es}).
  Fermion chiral $(L,R)$ indices are suppressed.  \label{tab:4ferm}}
\end{table}

The SMEFT Lagrangian is the sum of the dimension-4 terms and operators
defined in Tables~\ref{tab:no4ferm} and~\ref{tab:4ferm}:
\be
\lcal = \lcal_{\mathrm SM}^{(4)} + \frac{1}{\Lambda} C^{\nu\nu}
Q_{\nu\nu}^{(5)} + \frac{1}{\Lambda^2} \sum_{X} C^{X} Q_X^{(6)} +
\frac{1}{\Lambda^2} \sum_{f} C^{\prime f} Q_f^{(6)} \;.
\label{Leff}
\ee
Physical fields in SMEFT are obtained after the SSB.  We follow here
the systematic presentation (and notation) of
ref.~\cite{Dedes:2017zog}.  In the gauge and Higgs sectors physical
and Goldstone fields ($h, G^0,G^\pm, W^\pm_\mu, Z^0_\mu, A_\mu$) are
related to initial (Warsaw basis) fields ($\varphi, W_\mu^i, B_\mu,
G_\mu^A$) by the normalisation constants:
\bea
\left( \begin{array}{c} \varphi^+ \\ \varphi^0 \end{array} \right) &=&
\left ( \begin{array}{c} Z_{G^+}^{-1} G^+ \\ \frac{1}{\sqrt{2}} (v +
  Z_h^{-1} h + i Z_{G^0}^{-1} G^0) \end{array} \right ) \;,\nn
\left( \begin{array}{c} {B}_\mu \\ {W}^3_\mu \end{array} \right) & =&
\hat Z_{AZ}^{-1} \left (\begin{array}{c} {A}_\mu \\ {Z}_\mu
  \end{array} \right )  \;,
\label{eq:rot}\nn
W^1_{\mu} &=& {Z_W^{-1}\over \sqrt{2}}\,( W_\mu^+ + W_\mu^-)
\;, \label{Wpm-rot}\nn
W^2_{\mu} &=& {i Z_W^{-1}\over \sqrt{2}}\,( W_\mu^+ - W_\mu^-)
\;, \label{Wpm-rot1}\nn
G_\mu^A &=& Z_{G}^{-1} \, g_\mu^A \;.
\label{eq:hgnorm}
\eea
In addition, Feynman rules for physical fields are expressed in terms
of effective gauge couplings, chosen to preserve the natural form of
covariant derivative:
\bea
\begin{array}{l}
g\, = Z_{g} \bar g \qquad
g' = Z_{g'} \bar g' \qquad
g_s = Z_{g_s} \bar g_s\,.
\end{array}
\label{Zg-norm} 
\eea
In $d=6$ SMEFT $SU(2)$ and $SU(3)$ gauge field and gauge normalisation
constants are equal, $Z_g = Z_W$, $Z_{g_s} = Z_G$.  In addition
$Z_{g'} = 1 - v^2 C^{\varphi B}$ and $Z_{G^+}=1$.  Complete
expressions for the field normalisation constants, $Z_X$, for the
corrected Higgs field VEV, $v$, and for the gauge and Higgs boson
masses, $M_Z$, $M_W$ and $M_h$, including corrections from the full
set of dimension-5 and -6 operators, are given in
ref.~\cite{Dedes:2017zog}.  \sfr recalculates these corrections by
taking into account only the subset of non-vanishing SMEFT Wilson
coefficients chosen by the user, as described in Sec.~\ref{sec:sfr}.

The basis in the fermion sector is not fixed by the structure of gauge
interactions and allows for unitary rotation freedom in the flavour
space:
\begin{equation}
\psi_X^\prime = U_{\psi_X} \: \psi_X\;,
\end{equation}
with $\psi=\nu,e,u,d$ and $X=L,R$.  We choose the rotations such that
$\psi_X$ eigenstates correspond to real and non-negative eigenvalues
of $3\times 3$ fermion mass matrices:
\begin{equation}
\begin{array}{cc}
M^\prime_\nu = - v^2 C^{\prime \nu\nu}\;, &
M^\prime_e = \frac{v}{\sqrt{2}}\, \left (\Gamma_e - \frac{v^2}{2}
C^{\prime e\varphi} \right ),\; \\[2mm]
M^\prime_u = \frac{v}{\sqrt{2}}\, \left (\Gamma_u - \frac{v^2}{2}
C^{\prime u\varphi} \right ),\; &
M^\prime_d = \frac{v}{\sqrt{2}}\, \left (\Gamma_d - \frac{v^2}{2}
C^{\prime d\varphi} \right ).
\end{array}
\label{eq:modyuk}
\end{equation}
The fermion flavour rotations can be adsorbed in redefinitions of
Wilson coefficients, as a trace leaving CKM and PMNS matrices (denoted
in \sfr as $K$ and $U$ respectively) multiplying them.  It is
important to note a departure from~\Ref{Dedes:2017zog} in the
definition of these matrices, assumed here to be unitary products of
the form\footnote{In ref.~\cite{Dedes:2017zog}, these matrices are
defined as non-unitary ones, $K=U_{u_L}^\dagger U_{d_L}(1+ C^{\varphi
q (3)})$ and $U= (1+ C^{\varphi l(3)})U_{e_L}^\dagger U_{\nu_L}$.  It
is however more natural and economical to use the unitary definitions
of eq.~\eqref{eq:kudef} instead. The conventions for $K$ and $U$ of
ref.~\cite{Dedes:2017zog} can be still enforced in \sfr by editing the
file {\tt code/smeft\_variables.m} and changing there the value of
control variable {\tt SMEFT\$Unitary=0} to {\tt SMEFT\$Unitary=1}.}:
\bea
K = U_{u_L}^\dagger \, U_{d_L} \;, \qquad 
U = U_{e_L}^\dagger \, U_{\nu_L} \, .
\label{eq:kudef}
\eea
In practice, comparing to results of ref.~\cite{Dedes:2017zog} the
altered definition of eq.~(\ref{eq:kudef}) affects only four vertices,
namely couplings of $W$ boson and charged Goldstone to fermions:
$W^+(G^+)$-$\bar u^I$-$d^J$, $W^+(G^+)$-$\bar \nu^I$-$e^J$ and their
Hermitian conjugates.

The complete list of redefinitions of flavour-dependent Wilson
coefficients is given in Table~4 of ref.~\cite{Dedes:2017zog}.  After
rotations, they are defined in so called ``Warsaw mass'' basis (as
also described in WCxf standard~\cite{Aebischer:2017ugx}).  \sfr
assumes that the numerical values of Wilson coefficients are given in
this particular basis.

In summary, Feynman rules generated by the \sfr package describe
interactions of SMEFT physical (mass eigenstates) fields, with
numerical values of Wilson coefficients defined in the same (``Warsaw
mass'') basis.

It is also important to stress that in the general case of lepton
number flavour violation, with non-vanishing Weinberg operator
of~\eq{qnunu}, neutrinos are massive Majorana spinors, whereas under
the assumption of $L$-conservation they can be regarded as massless
Weyl spinors.  As described in the next Section, \sfr is capable to
generate Feynman rules for neutrino interactions in both cases,
depending on the choice of initial options.  One should remember that
treating neutrinos as Majorana particles requires special set of rules
for propagators, vertices and diagram combinatorics.  We follow here
the treatment of refs.~\cite{Denner:1992vza, Denner:1992me,
  Dedes:2017zog, Paraskevas:2018mks}.

\section{Deriving SMEFT Feynman rules  with \sfr package}
\label{sec:sfr}

\definecolor{lightgray}{rgb}{0.95,0.95,0.95}

\subsection{Installation}
\label{sec:install}

\sfr package works using the \frules system, so both need to be
properly installed first.  A recent version and installation
instructions for the \frules package can be downloaded from the
address:
\begin{center}
\url{https://feynrules.irmp.ucl.ac.be}
\end{center}
\sfr has been tested with \frules version 2.3.

Standard \frules installation assumes that the new models description
is put into {\tt Model} subdirectory of its main tree.  We follow this
convention, so that \sfr archive should be unpacked into
\begin{center}
 {\tt Models/SMEFT\_N\_NN }
\end{center}
catalogue, where {\tt N\_NN} denotes the package version (currently
version 2.00).  After installation, {\tt Models/SMEFT\_N\_NN }
contains the following files and subdirectories listed in
Table~\ref{tab:filestruct}.
  
\begin{table}[htb!]
\begin{mdframed}[backgroundcolor=lightgray]  
\begin{tabular}{p{43mm}p{100mm}}
{{\tt SmeftFR-init.nb smeft\_fr\_init.m }} & Notebook and equivalent
text script generating SMEFT Lagrangian in mass basis and Feynman
rules in {\em Mathematica} format. \\[2mm]
{{\tt SmeftFR-interfaces.nb smeft\_fr\_interfaces.m}} & Notebook and
text script with routines for exporting Feynman rules in various
formats: WCxf, Latex, UFO and FeynArts.\\[2mm]
{\tt SmeftFR.pdf} & package manual in pdf format.\\[2mm]
{\tt code} & subdirectory with package code and utilities.\\[2mm]
{\tt lagrangian} & subdirectory with expressions for the SM Lagrangian
and dimension-5 and 6 operators coded in \frules format.\\[2mm]
{\tt definitions} & subdirectory with templates of SMEFT ``model
files'' and example of numerical input for Wilson coefficients in WCxf
format.\\[2mm]
{\tt output} & subdirectory with dynamically generated ``parameter
files'' and output for Feynman rules in various formats, by default
{\em Mathematica}, Latex, UFO and FeynArts are generated.\\[2mm]
{\tt full\_rxi\_results} & subdirectory with ready-to-use set of
Feynman rules including all SMEFT operator classes, calculated in
$R_\xi$-gauges.\\[2mm]
{\tt full\_unitary\_results} & subdirectory with ready-to-use set of
Feynman rules including all SMEFT operator classes, calculated in
unitary gauge.
\end{tabular}
\end{mdframed}
\caption{Files and directories included in \sfr v2.00
  package. \label{tab:filestruct}}
\end{table}

Before running the package, one needs to set properly the main \frules
installation directory, defining the {\tt \$FeynRulesPath} variable at
the beginning of {\tt smeft\_init.m} and {\tt smeft\_outputs.m} files.
For non-standard installations (not advised!), also the variable {\tt
  SMEFT\$Path} has to be updated accordingly.

\subsection{Code structure}
\label{sec:code}

The most general version of SMEFT, including all possible flavour
violating couplings, is very complicated.  Symbolic operations on the
full SMEFT Lagrangian, including complete set of dimension 5 and 6
operators and with numerical values of all Wilson coefficients
assigned are time consuming and can take hours or even days on a
standard personal computer.  For most of the physical applications it
is sufficient to derive interactions only for a subset of
operators.\footnote{Eventually, operators must be selected with care
  as in general they may mix under
  renormalisation~\cite{Jenkins:2013zja, Jenkins:2013wua,
    Alonso:2013hga}.}

\begin{figure}[htb!]
\includegraphics[width=\textwidth,height=0.9\textheight]{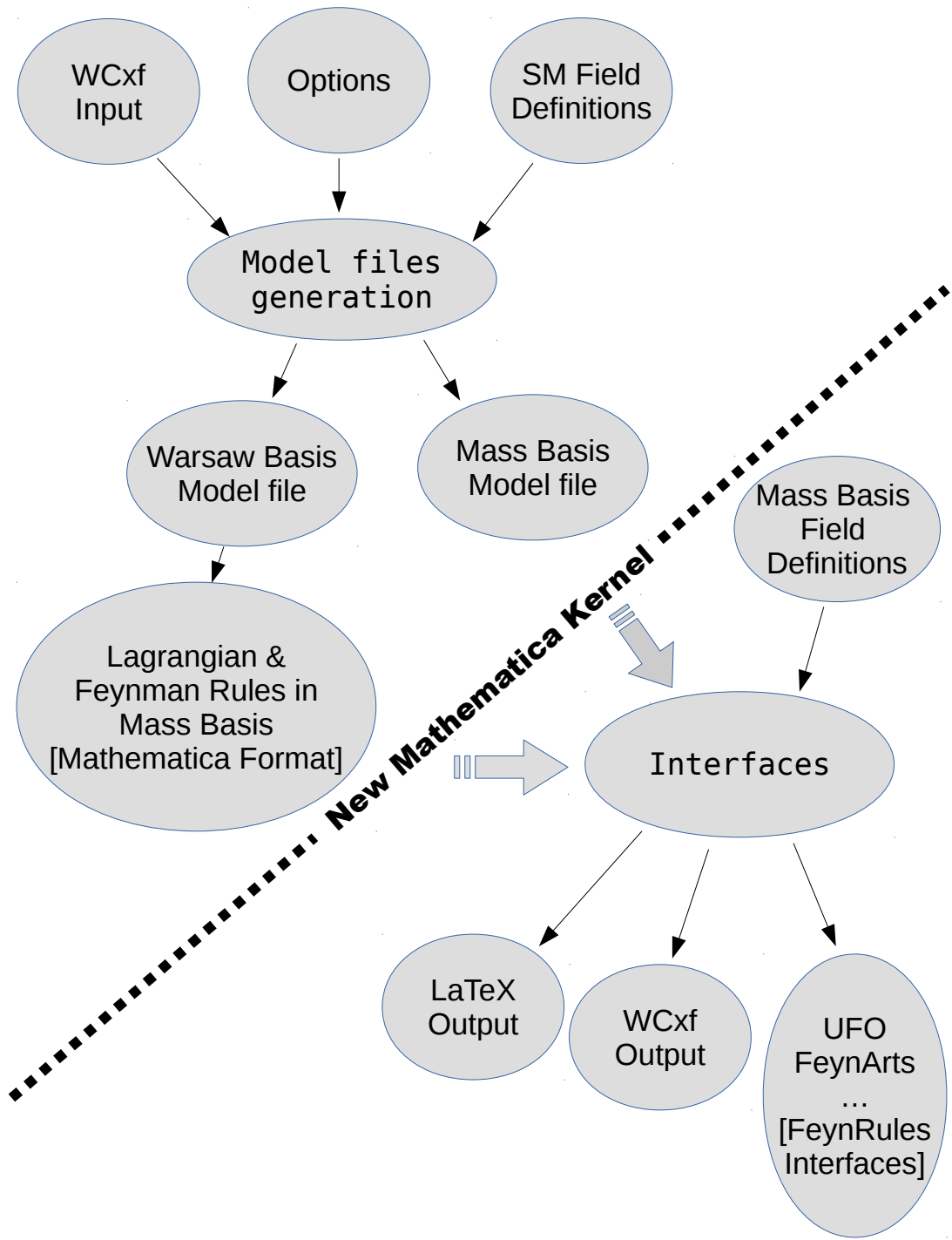}
  \caption{Structure of the \sfr code.  \label{fig:flow}}
\end{figure} 

To speed up the calculations, \sfr can evaluate Feynman rules for a
chosen subset of operators only, generating dynamically the proper
\frules ``model files''.  The calculations are divided in two stages,
as illustrated in flowchart of Fig.~\ref{fig:flow}.  First, the SMEFT
Lagrangian is initialised in Warsaw basis and transformed to mass
eigenstates basis analytically, truncating all terms of the order
${\cal O}(\frac{1}{\Lambda^3})$ and higher.  To speed up the program,
at this stage all flavour parameters are considered to be tensors with
indices without assigned numerical values (they are ``Internal''
parameters in \frules notation).  The resulting mass basis Lagrangian
and Feynman rules written in Mathematica format are stored on disk.
In the second stage, the previously generated output can be used
together with new ``model file'', this time containing numerical
values of (``External'') parameters, to export mass basis SMEFT
interactions in various commonly used external formats such as Latex,
WCxf and standard \frules supported interfaces -- UFO, FeynArts and
others.

\subsection{Model initialisation}
\label{sec:init}

\begin{table}[htb!]
\begin{mdframed}[backgroundcolor=lightgray]  
\begin{tabular}{lp{28mm}p{75mm}}
Option & Allowed values & Description \\[2mm]
\hline
\\
Operators & {default: all operators} & List with subset of SMEFT
operators included in calculations.  \\[3mm]
Gauge & {\bf Unitary}, Rxi & Choice of gauge fixing conditions \\[3mm]
WCXFInitFile & {\bf ""} & Name of file with numerical values of Wilson
coefficients in the WCxf format.  If this option is not set or the
file does not exist, all Wilson coefficients are set to $0$.\\[3mm]
MajoranaNeutrino & {\bf False}, True & Neutrino fields are treated as
Majorana spinors if $Q_{\nu\nu}$ is included in the operator list,
massless Weyl spinors otherwise.  Setting this option to {\bf True}
allows one to use Majorana spinors also in the massless case.  \\[3mm]
Correct4Fermion & False, {\bf True} & Corrects relative sign of some
4-fermion interactions, fixing results produced by \frules.  \\[3mm]
WBFirstLetter & {\bf "c"} & Customizable first letter of Wilson
coefficient names in Warsaw basis (default $c_G, \ldots$). Can be used
to avoid convention clashes when comparing with other SMEFT bases.
\\[3mm]
MBFirstLetter & {\bf "C"} & Customizable first letter of Wilson
coefficient names in mass basis (default $C_G, \ldots$).
\end{tabular}
\end{mdframed}  
\caption{The allowed options of {\tt SMEFTInitializeModel} routine.
  If an option is not specified, the default value (marked above in
  boldface) is assumed.
\label{tab:init}
}
\end{table}

In the first step, the relevant \frules model files must be generated.
This is done by calling the function:

\bigskip

{\tt SMEFTInitializeModel}[{\it Option1 $\to$ Value1, Option2 $\to$
    Value2, $\ldots$}]

\bigskip

\noindent with the allowed options listed in Table~\ref{tab:init}.

Names of operators used in \sfr are derived from the subscript indices
of operators listed in Tables~\ref{tab:no4ferm} and~\ref{tab:4ferm},
with obvious transcriptions of ``tilde'' symbol and Greek letters to
Latin alphabet.  By default, all possible 59+1+4 SMEFT ($d=6$)
operator classes are included in calculations, which is equivalent to
setting:

\bigskip

\noindent {\tt Operators} $\to$ \{ {\tt "G"}, {\tt "Gtilde"}, {\tt
  "W"}, {\tt "Wtilde"}, {\tt "phi"}, {\tt "phiBox"}, {\tt "phiD"},
          {\tt "phiW"}, {\tt "phiB"}, {\tt "phiWB"}, {\tt
            "phiWtilde"}, {\tt "phiBtilde"}, {\tt "phiWtildeB"}, {\tt
            "phiGtilde"}, {\tt "phiG"}, {\tt "ephi"}, {\tt "dphi"},
          {\tt "uphi"}, {\tt "eW"}, {\tt "eB"}, {\tt "uG"}, {\tt
            "uW"}, {\tt "uB"}, {\tt "dG"}, {\tt "dW"}, {\tt "dB"},
          {\tt "phil1"}, {\tt "phil3"}, {\tt "phie"}, {\tt "phiq1"},
          {\tt "phiq3"}, {\tt "phiu"}, {\tt "phid"}, {\tt "phiud"},
          {\tt "ll"}, {\tt "qq1"}, {\tt "qq3"}, {\tt "lq1"}, {\tt
            "lq3"}, {\tt "ee"}, {\tt "uu"}, {\tt "dd"}, {\tt "eu"},
          {\tt "ed"}, {\tt "ud1"}, {\tt "ud8"}, {\tt "le"}, {\tt
            "lu"}, {\tt "ld"}, {\tt "qe"}, {\tt "qu1"}, {\tt "qu8"},
          {\tt "qd1"}, {\tt "qd8"}, {\tt "ledq"}, {\tt "quqd1"}, {\tt
            "quqd8"}, {\tt "lequ1"}, {\tt "lequ3"}, {\tt "vv"}, {\tt
            "duq"}, {\tt "qqu"}, {\tt "qqq"}, {\tt "duu"} \}

\bigskip

To speed up the derivation of Feynman rules and to get more compact
expressions, the user can restrict the list above to any preferred
subset of operators.

\sfr is fully integrated with the WCxf standard.  Apart from
numerically editing Wilson coefficients in \frules model files,
reading them from the WCxf input is the only way of automatic
initialisation of their numerical values.  Such an input format is
exchangeable between a larger set of SMEFT-related public
packages~\cite{Aebischer:2017ugx} and may help to compare their
results.

An additional advantage of using WCxf input format comes in the
flavour sector of the theory.  Here, Wilson coefficients are in
general tensors with flavour indices, in many cases symmetric under
various permutations.  WCxf input requires initialisation of only the
minimal set of flavour dependent Wilson coefficients, those which
could be derived by permutations are also automatically properly
set.\footnote{We would like to thank D.~Straub for supplying us with a
  code for symmetrisation of flavour-dependent Wilson coefficients.}

Further comments concern {\tt MajoranaNeutrino} and {\tt
  Correct4Fermion} options.  They are used to modify the analytical
expressions only for the Feynman rules, not at the level of the mass
basis Lagrangian from which the rules are derived.  This is because
some \frules interfaces, like UFO, intentionally leave the relative
sign of 4-fermion interactions uncorrected\footnote{B.~Fuks, private
  communication.}, as it is later changed by Monte Carlo generators
like MadGraph5.  Correcting the sign before generating UFO output
would therefore lead to wrong final result.  Similarly, treatment of
neutrinos as Majorana fields could not be compatible with hard coded
quantum number definitions in various packages.  On the other hand, in
the manual or symbolic computations it is convenient to have from the
start the correct form of Feynman rules, as done by \sfr when both
options are set to their default values.

{\tt SMEFTInitializeModel} routine does not require prior loading of
\frules package.  After execution, it creates in the {\tt output}
subdirectory three model files listed in Table~\ref{tab:modelfiles}.
Parameter files generated by {\tt SMEFTInitializeModel} contain also
definitions of SM parameters, copied from templates {\tt
  definitions/smeft\_par\_head\_WB.fr} and {\tt
  definitions/smeft\_par\_head\_MB.fr}.  The values of SM parameters
can be best updated directly by editing the template files and the
header of the {\tt code/smeft\_variables.m} file, otherwise they will
be overwritten in each rerun of \sfr initialisation routines.

\begin{table}[htb!]
\begin{center}
\begin{mdframed}[backgroundcolor=lightgray]  
\begin{tabular}{lp{95mm}}
{\tt smeft\_par\_WB.par} & SMEFT parameter file with Wilson
coefficients in Warsaw basis (defined as ``Internal'', with no
numerical values assigned).\\[1mm]
{\tt smeft\_par\_MB.par} & SMEFT parameter file with Wilson
coefficients in mass basis (defined as ``External'', numerical values
imported from the input file in WCxf format).\\[1mm]
{\tt smeft\_par\_MB\_real.par} & as {\tt smeft\_par\_MB.par}, but only
real values of Wilson coefficients given in WCxf file are included in
SMEFT parameter file, as required by many event generators.
\end{tabular}    
\end{mdframed}
\end{center}
\caption{Model files generated by the {\tt SMEFTInitializeModel}
  routine.\label{tab:modelfiles}}
\end{table}

As mentioned above, in all analytical calculations performed by \sfr,
terms suppressed by ${\cal O}(1/\Lambda^3)$ or higher $1/\Lambda$
powers are always neglected.  Therefore, the resulting Feynman rules
can be consistently used to calculate physical observables,
symbolically or numerically by Monte Carlo generators, up to the
linear order in dimension-6 operators.  This information is encoded in
\frules SMEFT model files by assigning the ``interaction order''
parameter {\tt NP=1} to each Wilson coefficients and setting in {\tt
  smeft\_field\_WB.fr} and {\tt smeft\_field\_MB.fr} the limits:

\bigskip
\begin{tabular}{l}
M\$InteractionOrderLimit = \{ \\
~~~ \{QCD,99\}, \\
~~~ \{NP,1\}, \\
~~~ \{QED,99\} \\
~ \} 
\end{tabular}
\bigskip

\noindent If one wishes to include (e.g. for testing purposes) terms
quadratic in Wilson coefficients in the automatic cross section
calculations performed by matrix element and Monte Carlo generators,
one should edit both model files and increase the allowed order of NP
contributions by changing the definition of {\tt
  M\$InteractionOrderLimit}, setting there {\tt \{NP,2\}}.

An additional remark concerns the value of neutrino masses.  In mass
basis, the neutrino masses are equal to $-v^2 C_{\nu\nu}^{II}$~ [see
  \eq{eq:modyuk}].  Thus, the numerical values of $C_{\nu\nu}^{II}$
coefficients should be real and negative.  If positive or complex
values of $C_{\nu\nu}^{II}$ are given in the WCxf input file, then the
{\tt SMEFTInitializeModel} routine evaluates neutrino masses as
$M_{\nu^I} = v^2 |C_{\nu\nu}^{II}|$.

\subsection{Calculation of mass basis Lagrangian and Feynman rules}
\label{sec:mbasis}

By loading the \frules model files the derivation of SMEFT Lagrangian
in mass basis is performed by calling the following sequence of
routines:
\begin{center}
%
%
\begin{tabular}{lp{95mm}}
{\tt SMEFTLoadModel[ ]} & Loads {\tt output/smeft\_par\_WB.par} model
file and calculates SMEFT Lagrangian in Warsaw basis for chosen subset
of operators \\
{\tt SMEFTFindMassBasis[ ]} & Finds field bilinears and analytical
transformations diagonalizing mass matrices up to ${\cal
  O}(1/\Lambda^2)$ \\[1mm]
{\tt SMEFTFeynmanRules[ ]} & Evaluates analytically SMEFT Lagrangian
and Feynman rules in the mass basis, again truncating consistently all
terms higher then ${\cal O}(1/\Lambda^2)$.
\end{tabular}
%
%
\end{center}

The calculation time may vary considerably depending on the choice of
operator (sub-)set and gauge fixing conditions chosen.  For the full
list of SMEFT $d=5$ and $d=6$ operators and in $R_\xi$-gauges, one can
expect CPU time necessary to evaluate all Feynman rules, from about an
hour to many hours on a typical personal computer, depending on its
speed capabilities.

One should note that when neutrinos are treated as Majorana particles,
(as necessary in case of non-vanishing Wilson coefficient of $d=5$
Weinberg operator), their interactions involve lepton number
non-conservation.  When \frules is dealing with them it produces
warnings of the form:

\bigskip
\noindent \red{{\it QN::NonConserv: Warning: non quantum number
    conserving vertex encountered!}\\[1mm]
{\it Quantum number LeptonNumber not conserved in vertex $\ldots$}
}
\bigskip

\noindent Obviously such warnings should be ignored.

Evaluation of Feynman rules for vertices involving more than two
fermions is not fully implemented yet in {\tt FeynRules}. To our
experience, apart from the issue of relative sign of four fermion
diagrams mentioned earlier, particularly problematic was the correct
automatic derivation of quartic interactions with four Majorana
neutrinos and similar vertices which violate $B$- and $L$-quantum
numbers. For these special cases, \sfr overwrites the \frules result
with manually calculated formulae encoded in Mathematica format.

Another remark concerns the hermicity property of the SMEFT
Lagrangian. For some types of interactions, e.g. four-fermion vertices
involving two-quarks and two-leptons, the function {\tt
  CheckHermicity} provided by \frules reports non-Hermitian terms in
the Lagrangian. However, such terms are actually Hermitian if
permutation symmetries of indices of relevant Wilson coefficients are
taken into account.  Such symmetries are automatically imposed if
numerical values of Wilson coefficients are initialized with the use
of {\tt SMEFTInitializeMB} or {\tt SMEFTToWCXF} routines (see
Sections~\ref{sec:interface} and~\ref{sec:wcxf}).

Results of the calculations are collected in file {\tt
  output/smeft\_feynman\_rules.m}. The Feynman rules and pieces of the
mass basis Lagrangian for various classes of interactions are stored
in the variables with self-explanatory names listed in
Table~\ref{tab:variables}.

\begin{table}[tb!]
\begin{center}
\begin{mdframed}[backgroundcolor=lightgray]  
{\tt 
\begin{tabular}{lp{20mm}l}
  LeptonGaugeVertices &&   QuarkGaugeVertices \\
  LeptonHiggsGaugeVertices &&  QuarkHiggsGaugeVertices \\
  QuarkGluonVertices \\
  GaugeSelfVertices &&  GaugeHiggsVertices \\
  GluonSelfVertices &&  GluonHiggsVertices \\
  GhostVertices \\
  FourLeptonVertices &&   FourQuarkVertices \\
  TwoQuarkTwoLeptonVertices \\
  DeltaLTwoVertices && BLViolatingVertices 
\end{tabular}
}
\end{mdframed}
\end{center}
\caption{Names of variables defined in the file {\tt
    output/smeft\_feynman\_rules.m} containing expressions for Feynman
  rules.  Parts of mass basis Lagrangian are stored in equivalent set
  of variables, with ``{\tt Vertices}'' replaced by ``{\tt
    Lagrangian}'' in part of their names (i.e.  {\tt
    LeptonGaugeVertices} $\to$ {\tt LeptonGaugeLagrangian},
  \textit{etc.}). \label{tab:variables}}
\end{table}

File {\tt output/smeft\_feynman\_rules.m} contains also expressions
for the normalisation factors relating Higgs and gauge fields and
couplings in the Warsaw and mass basis. Namely, variables {\tt Hnorm,
  G0norm, GPnorm, AZnorm[i,j], Wnorm, Gnorm}, correspond to,
respectively, $Z_h^{-1}$, $Z_{G^0}^{-1}$, $Z_{G^+}^{-1}$ $\hat
Z_{AZ}^{-1}$, $Z_W^{-1}$ and $Z_G^{-1}$ in~\eqref{eq:hgnorm}.  In
addition, formulae for tree level corrections to SM mass parameters
and Yukawa couplings are stored in variables {\tt SMEFT\$vev}, {\tt
  SMEFT\$MH2}, {\tt SMEFT\$MW2}, {\tt SMEFT\$MZ2}, {\tt
  SMEFT\$YL[i,j]}, {\tt SMEFT\$YD[i,j]} and {\tt SMEFT\$YU[i,j]}.

It is important to note that although at this point the Feynman rules
for the mass basis Lagrangian are already calculated, definitions for
fields and parameters used to initialise the SMEFT model in \frules
are still given in Warsaw basis.  To avoid inconsistencies, it is
strongly advised to quit the current Mathematica kernel and start new
one reloading the mass basis Lagrangian together with the compatible
model files with fields defined also in mass basis, as described next
in Sec.~\ref{sec:interface}.  All further calculations should be
performed within this new kernel.

\subsection{Interfaces}
\label{sec:interface}

\sfr output in some of portable formats must be generated from the
SMEFT Lagrangian transformed to mass basis, with all numerical values
of parameters initialised.  As \frules does not allow for two
different model files loaded within a single \textit{Mathematica}
session, one needs to quit the kernel used to run routines necessary
to obtain Feynman rules and, as described in previous Section, start a
new \textit{Mathematica} kernel.  Within it, the user must reload
\frules and \sfr packages and call the following routine:

\bigskip

{\tt SMEFTInitializeMB[ {\it Options} ]}

\bigskip

\noindent Allowed options are given in Table~\ref{tab:mbinit}.  After
call to {\tt SMEFTInitializeMB}, mass basis model files are read and
the mass basis Lagrangian is stored in a global variable named {\tt
  SMEFTMBLagrangian} for further use by interface routines.

\begin{table}[tb!]
\begin{center}
\begin{mdframed}[backgroundcolor=lightgray]  
\noindent \begin{tabular}{lp{20mm}p{85mm}}
Option  & Allowed values & Description \\[2mm]
\hline
\\
RealParameters & {\bf False}, True & Default initialisation is done
using {\tt output/smeft\_par\_MB.par} file, which may contain complex
parameters, not compatible with matrix element generators.  Setting
{\it RealParameters $\to$ True} forces loading of {\tt
  output/smeft\_par\_MB\_real.par} file where imaginary parts of all
Wilson coefficients are set to $0$. Imaginary phases of CKM and PMNS
matrices, if present, are also set to zero after loading this
file.\\[3mm]
Include4Fermion & {\bf False}, True & 4-fermion vertices are not fully
implemented in \frules and by default not included in SMEFT
interactions.  Set this option to True to include such terms.
\end{tabular}
\end{mdframed}
\end{center}
\caption{Options of {\tt SMEFTInitializeMB} routine, with default
  values marked in boldface.\label{tab:mbinit}}
\end{table}

\subsubsection{WCxf input and output}
\label{sec:wcxf}

Translation between \frules model files and WCxf format is done by the
functions {\tt SMEFTToWCXF} and {\tt WCXFToSMEFT}. They can be used
standalone and do not require loading \frules and calling first {\tt
  SMEFTInitializeMB} routine to work properly.

Exporting numerical values of Wilson coefficients of operators in the
WCxf format is done by the function:

\bigskip

{\tt SMEFTToWCXF[ SMEFT\_Parameter\_File, WCXF\_File ] }

\bigskip

\noindent where the arguments {\tt SMEFT\_Parameter\_File, WCXF\_File}
define the input model parameter file in the \frules format and the
output file in the WCxf JSON format, respectively.  The created JSON
file can be used to transfer numerical values of Wilson coefficients
to other codes supporting WCxf format.
Note that in general, the \frules model files may contain different
classes of parameters, according to the {\tt Value} property defined
to be a number (real or complex), a formula or even not defined at
all. Only the Wilson coefficients with {\tt Value} defined to be a
number are transferred to the output file in WCxf format.

Conversely, files in WCxf format can be translated to \frules
parameter files using:

\bigskip

{\tt WCXFToSMEFT[ WCXF\_File, SMEFT\_Parameter\_File {\it Options}] }

\bigskip

\noindent with the allowed options defined in
Table~\ref{tab:wcxf2}.

\begin{table}[htb!]
\begin{mdframed}[backgroundcolor=lightgray]  
\begin{tabular}{lp{27mm}p{80mm}}
Option & Allowed values & Description \\[2mm]
\hline\\
Operators & default: all operators & List with subset of Wilson
coefficients to be included in the SMEFT parameter file \\[3mm]
RealParameters & False, {\bf True} & Decides if only real values of
Wilson coefficients given in WCxf file are included in SMEFT parameter
file \\[3mm]
OverwriteTarget & {\bf False}, True & If set to True, target file is
overwritten without warning \\[3mm]
Silent & {\bf False}, True & Debug option, suppresses screen comments
\\[3mm]
FirstLetter & {\bf "C"} & Customizable first letter of Wilson
coefficient names in mass basis (default $C_G, \ldots$).
\end{tabular}
\end{mdframed}  
\caption{Options of {\tt WCXFToSMEFT} routine.   Default values are
  marked in boldface.\label{tab:wcxf2}}
\end{table}

\subsubsection{Latex output}
\label{sec:latex}

\sfr provides a dedicated Latex generator (not using the generic
\frules Latex export routine).  Its output has the following
structure:
\begin{itemize}
\item For each interaction vertex, the diagram is drawn, using the
  {\tt axodraw} style~\cite{Vermaseren:1994je}.  Expressions for
  Feynman rules are displayed next to corresponding diagrams.
\item In analytical expressions, all terms multiplying a given Wilson
  coefficient are collected together and simplified.
\item Long analytical expressions are automatically broken into many
  lines using {\tt breakn} style (this does not always work perfectly
  but the printout is sufficiently readable).
\end{itemize}
Latex output is generated by the function:

\bigskip

{\tt SMEFTToLatex[ {\it Options} ] }

\bigskip

\noindent with the allowed options listed in Table~\ref{tab:latex}.
The function {\tt SMEFTToLatex} assumes that the variables listed in
Table~\ref{tab:variables} are initialised. It can be called either
after executing relevant commands, described in Sec.~\ref{sec:mbasis},
or after reloading the mass basis Lagrangian with the {\tt
  SMEFTInitializeMB} routine, see Sec.~\ref{sec:interface}.

\begin{table}[tb!]
\begin{center}
\begin{mdframed}[backgroundcolor=lightgray]  
\noindent \begin{tabular}{lp{27mm}p{85mm}}
Option name & Allowed values & Description \\[2mm]
\hline\\
FullDocument & False, {\bf True} & By default a complete document is
generated, with all headers necessary for compilation.  If set to
False, headers are stripped off and the output file can, without
modifications, be included into other Latex documents.  \\[3mm]
ScreenOutput & {\bf False}, True & For debugging purposes, if set to
True the Latex output is printed also to the screen.
\end{tabular}
\end{mdframed}
\end{center}
\caption{Options of {\tt SMEFTToLatex} routine, with default values
  marked in boldface.\label{tab:latex}}
\end{table}

Latex output is stored in {\tt output/latex subdirectory}, split into
smaller files each containing one primary vertex.  The main file is
named {\tt smeft\_feynman\_rules.tex}.  The style files necessary to
compile Latex output are supplied with the \sfr distribution.

Note that the correct compilation of documents using ``axodraw.sty''
style requires creating intermediate Postscript file.  Programs like
{\it pdflatex} producing directly PDF output will not work properly.
One should instead use e.g.:

\bigskip

{\tt

  latex  smeft\_feynman\_rules.tex

  dvips smeft\_feynman\_rules.dvi

  ps2pdf smeft\_feynman\_rules.ps
}

\bigskip

The {\tt smeft\_feynman\_rules.tex} does not contain analytical
expressions for five and six gluon vertices.  Such formulae are very
long (multiple pages, hard to even compile properly) and not useful
for hand-made calculations.  If such vertices are needed, they should
be rather directly exported in some other formats as described in the
next subsection.

Other details not printed in the Latex output, such as, the form of
field propagators, conventions for parameters and momenta flow in
vertices (always incoming), manipulation of four-fermion vertices with
Majorana fermions \textit{etc}, are explained thoroughly in the
Appendices A1--A3 of ref.~\cite{Dedes:2017zog}.

\subsubsection{Standard \frules interfaces}
\label{sec:ufo}

After calling the initialisation routine {\tt SMEFTInitializeMB}, the
output to UFO, FeynArts and other formats supported by \frules
interfaces, can be generated using \frules commands and options from
the mass basis Lagrangian stored in the {\tt SMEFTMBLagrangian}
variable.  For instance, one could call:

\bigskip

\noindent {\tt WriteUFO[ SMEFTMBLagrangian, {\it Output $\to$
      "output/UFO", AddDecays $\to$ False, \ldots}] }

\noindent {\tt WriteFeynArtsOutput[ SMEFTMBLagrangian, {\it Output
      $\to$ "output/FeynArts", \ldots}] }

\bigskip

\noindent and similarly for other formats.

It is important to note that \frules interfaces like UFO or FeynArts
generate their output starting from the level of SMEFT mass basis
Lagrangian.  Thus, options of {\tt SMEFTInitializeModel} function like
{\tt MajoranaNeutrino} and {\tt Correct4Fermion} (see
Table~\ref{tab:init}) have no effect on output generated by the
interface routines.  As explained in Section~\ref{sec:init} they
affect only the expressions for Feynman rules.

If four-fermion vertices are included in SMEFT Lagrangian, UFO
produces warning messages of the form:

\bigskip
\red{{\it Warning: Multi-Fermion operators are not yet fully
    supported!}}
\bigskip

Therefore, the output for four-fermion interactions in UFO or other
formats must be treated with care and limited trust --- performing
appropriate checks are left to users' responsibility.  To our
experience, implementation in \frules of baryon and lepton number
violating four-fermion interactions, with charge conjugation matrix
appearing explicitly in vertices, is even more problematic.  Thus, for
safety in current \sfr version (2.00) such terms are never included in
{\tt SMEFTMBLagrangian} variable, eventually they can be passed to
interface routines separately via the {\tt BLViolatingLagrangian}
variable.

Exporting to UFO or other formats can take a long time, even several
hours for $R_\xi$-gauges and complete SMEFT Lagrangian with fully
general flavour structure and all numerical values of parameters
initialised.

Finally, it is important to stress here that our Feynman rules
communicate properly with MadGraph5 and FeynArts.  In particular, we
ran without errors test simulations in MadGraph5 using UFO model files
produced by \sfr v2. Similar tests were performed with amplitude
generation for sample processes using \sfr v2 FeynArts output.

\section{Sample programs}
\label{sec:sample}

After setting the variable {\tt \$FeynRulesPath} to correct value, in
order to evaluate mass basis SMEFT Lagrangian and analytical form of
Feynman rules one can use the following sequence of commands:

\bigskip

{\tt

\noindent SMEFT\$MajorVersion      = "2";\\
SMEFT\$MinorVersion      = "00";\\
SMEFT\$Path              = FileNameJoin[\{\$FeynRulesPath, "Models", "SMEFT\_" <> \\
\hspace*{4cm} SMEFT\$MajorVersion <> "\_" <> SMEFT\$MinorVersion\}];\\
\\
Get[ FileNameJoin[{\$FeynRulesPath,"FeynRules.m"}] ];\\
Get[ FileNameJoin[{ SMEFT\$Path, "code", "smeft\_package.m"}] ];\\
\\
OpList = \{ } {\tt "G"}, {\tt "Gtilde"}, {\tt "W"}, {\tt "Wtilde"},
{\tt "phi"}, {\tt "phiBox"}, {\tt "phiD"}, {\tt "phiW"}, {\tt "phiB"},
{\tt "phiWB"}, {\tt "phiWtilde"}, {\tt "phiBtilde"}, {\tt
  "phiWtildeB"}, {\tt "phiGtilde"}, {\tt "phiG"}, {\tt "ephi"}, {\tt
  "dphi"}, {\tt "uphi"}, {\tt "eW"}, {\tt "eB"}, {\tt "uG"}, {\tt
  "uW"}, {\tt "uB"}, {\tt "dG"}, {\tt "dW"}, {\tt "dB"}, {\tt
  "phil1"}, {\tt "phil3"}, {\tt "phie"}, {\tt "phiq1"}, {\tt "phiq3"},
{\tt "phiu"}, {\tt "phid"}, {\tt "phiud"}, {\tt "ll"}, {\tt "qq1"},
{\tt "qq3"}, {\tt "lq1"}, {\tt "lq3"}, {\tt "ee"}, {\tt "uu"}, {\tt
  "dd"}, {\tt "eu"}, {\tt "ed"}, {\tt "ud1"}, {\tt "ud8"}, {\tt "le"},
{\tt "lu"}, {\tt "ld"}, {\tt "qe"}, {\tt "qu1"}, {\tt "qu8"}, {\tt
  "qd1"}, {\tt "qd8"}, {\tt "ledq"}, {\tt "quqd1"}, {\tt "quqd8"},
{\tt "lequ1"}, {\tt "lequ3"}, {\tt "vv"}, {\tt "duq"}, {\tt "qqu"},
{\tt "qqq"}, {\tt "duu"} {\tt \};\\
\\  
SMEFTInitializeModel[ Operators -> OpList, Gauge -> Rxi, WCXFInitFile
  -> "wcxf\_input\_file\_with\_path.json" ];\\
\\
SMEFTLoadModel[ ];\\
SMEFTFindMassBasis[ ];\\
SMEFTFeynmanRules[ ];

}

\bigskip

\noindent or alternatively rerun the supplied programs: the notebook
  {\tt SmeftFR-init.nb} or the text script {\tt smeft\_fr\_init.m}.

As described before, Latex, WCxf, UFO and FeynArts formats can be
exported after rerunning first {\tt SmeftFR-init.nb} or equivalent set
of commands generating file {\tt smeft\_feynman\_rules.m} containing
the expressions for the mass basis Lagrangian.  Then, the user needs
to start a new \textit{Mathematica} kernel and rerun the notebook file
{\tt SmeftFR-interfaces.nb} or the script {\tt
  smeft\_fr\_interfaces.m}.  Alternatively, one can manually type the
commands, if necessary changing some of their options as described in
previous sections:

\bigskip

{\tt
\noindent Get[ FileNameJoin[\{\$FeynRulesPath,"FeynRules.m"\}] ];\\
Get[ FileNameJoin[\{SMEFT\$Path, "code", "smeft\_package.m"\}] ];\\
\\
SMEFTInitializeMB[ ];\\
\\
SMEFTToWCXF[ FileNameJoin[\{SMEFT\$Path, "output",
    "smeft\_par\_MB.fr"\}], \\
\hspace*{26mm} FileNameJoin[\{SMEFT\$Path, "output",
      "smeft\_wcxf\_MB.json"\}] ];\\
SMEFToLatex[ ];\\
\\
WriteUFO[ SMEFTMBLagrangian, $\ldots$ ];\\
WriteFeynArtsOutput[ SMEFTMBLagrangian, $\ldots$ ];\\
}

Feynman rules calculated for the full SMEFT operator set and two
possible gauge choices (unitary and $R_\xi$), including also outputs
in Latex, UFO and FeynArts formats, are stored in the subdirectories
{\tt full\_rxi\_results} and {\tt full\_unitary\_results}.  They can
be directly used, without rerunning the \sfr package.  However, such a
general output with all operators and numerical values for all 2499
SMEFT parameters initialised, is huge and importing it directly to
other SMEFT codes may cause them to work very slowly.

\section{Summary}
\setcounter{equation}{0}
\label{sec:summary}

The proliferation of the primitive vertices in SMEFT, even in
non-redundant basis of effective field operators, suggests a certain
kind of computational aid.  During the last few years, such an effort
has been intensive between high energy physicists.  Aiming at this
direction, we present here a code, named {\tt SmeftFR}, that generates
Feynman rules in SMEFT with $\text{dimension}\le 6$ operators given in
``Warsaw basis''~\cite{Grzadkowski:2010es} without restricting to
specific flavour structure, $CP$-, $B$- nor $L$-number conservation.
We have exploited the quantisation steps of SMEFT in unitary and
$R_\xi$-gauges following a procedure described in
ref.~\cite{Dedes:2017zog}.  {\tt SmeftFR} has been written on top of
the \textit{Mathematica} package {\tt FeynRules}.

In this article, we describe how to use {\tt SmeftFR} package in order
to produce Feynman rules for a selection of operators relevant to
observable (or observables) under study, with further options to
handle massive Majorana or massless Weyl neutrinos.  The output of the
package can be printed in Latex or exported in various formats
supported by {\tt FeynRules}, such as UFO, FeynArts, {\it etc}.  Input
parameters for Wilson coefficients used in {\tt SmeftFR} can
communicate with WCxf format for further numerical handling.  Feynman
rules are given in SMEFT mass basis and in both unitary or linear
$R_\xi$-gauges for further computational checks.

The current version of \sfr code and its manual can be downloaded from
the address
\begin{center}
\webpage
\end{center}

We believe that {\tt SmeftFR} is bridging a gap between the effective
SM Lagrangian all the way down to amplitude calculations required by
the experimental analyses.

\section*{Acknowledgements}

The work of JR was supported in part by Polish National Science Centre
under research grants
DEC-2015/19/B/ST2/02848,
and DEC-2015/18/M/ST2/00054.
MP acknowledges financial support by the Polish National Science
Center under the Beethoven series grant number DEC-2016/23/G/ST2/04301.
KS would like to thank the Greek State Scholarships Foundation (IKY)
for full financial support through the Operational Programme ``Human
Resources Development, Education and Lifelong Learning, 2014-2020''.
LT is supported by the Onassis Foundation --- Scholarship ID: G ZO
029-1/2018-2019.
JR would like to thank University of Ioannina and CERN for the
hospitality during his stays there.



\bibliographystyle{elsarticle-num}
\bibliography{EFT}{}

\begin{thebibliography}{10}
\expandafter\ifx\csname url\endcsname\relax
  \def\url#1{\texttt{#1}}\fi
\expandafter\ifx\csname urlprefix\endcsname\relax\def\urlprefix{URL }\fi
\expandafter\ifx\csname href\endcsname\relax
  \def\href#1#2{#2} \def\path#1{#1}\fi

\bibitem{Weinberg:1967tq}
S.~Weinberg, {A Model of Leptons}, Phys.Rev.Lett. 19 (1967) 1264--1266.
\newblock \href {https://doi.org/10.1103/PhysRevLett.19.1264}
  {\path{doi:10.1103/PhysRevLett.19.1264}}.

\bibitem{Glashow}
S.~Glashow, {Partial Symmetries of Weak Interactions}, Nucl.Phys. 22 (1961)
  579--588.
\newblock \href {https://doi.org/10.1016/0029-5582(61)90469-2}
  {\path{doi:10.1016/0029-5582(61)90469-2}}.

\bibitem{Salam}
A.~SalamIn {\it Proceedings of the Eighth Nobel Symposium}, edited by N.
  Svartholm (Wiley, New York, 1968), p.367.

\bibitem{Weinberg:1980wa}
S.~Weinberg, {Effective Gauge Theories}, Phys. Lett. B91 (1980) 51--55.
\newblock \href {https://doi.org/10.1016/0370-2693(80)90660-7}
  {\path{doi:10.1016/0370-2693(80)90660-7}}.

\bibitem{Coleman:1969sm}
S.~R. Coleman, J.~Wess, B.~Zumino, {Structure of phenomenological Lagrangians.
  1.}, Phys. Rev. 177 (1969) 2239--2247.
\newblock \href {https://doi.org/10.1103/PhysRev.177.2239}
  {\path{doi:10.1103/PhysRev.177.2239}}.

\bibitem{Callan:1969sn}
C.~G. Callan, Jr., S.~R. Coleman, J.~Wess, B.~Zumino, {Structure of
  phenomenological Lagrangians. 2.}, Phys. Rev. 177 (1969) 2247--2250.
\newblock \href {https://doi.org/10.1103/PhysRev.177.2247}
  {\path{doi:10.1103/PhysRev.177.2247}}.

\bibitem{Manohar:2018aog}
A.~V. Manohar, {Introduction to Effective Field Theories}, in: {Les Houches
  summer school: EFT in Particle Physics and Cosmology Les Houches, Chamonix
  Valley, France, July 3-28, 2017}, 2018.
\newblock \href {http://arxiv.org/abs/1804.05863} {\path{arXiv:1804.05863}}.

\bibitem{Buchmuller:1985jz}
W.~Buchmuller, D.~Wyler, {Effective Lagrangian Analysis of New Interactions and
  Flavor Conservation}, Nucl. Phys. B268 (1986) 621--653.
\newblock \href {https://doi.org/10.1016/0550-3213(86)90262-2}
  {\path{doi:10.1016/0550-3213(86)90262-2}}.

\bibitem{Grzadkowski:2010es}
B.~Grzadkowski, M.~Iskrzynski, M.~Misiak, J.~Rosiek, {Dimension-Six Terms in
  the Standard Model Lagrangian}, JHEP 10 (2010) 085.
\newblock \href {http://arxiv.org/abs/1008.4884} {\path{arXiv:1008.4884}},
  \href {https://doi.org/10.1007/JHEP10(2010)085}
  {\path{doi:10.1007/JHEP10(2010)085}}.

\bibitem{Dedes:2017zog}
A.~Dedes, W.~Materkowska, M.~Paraskevas, J.~Rosiek, K.~Suxho, {Feynman rules
  for the Standard Model Effective Field Theory in R$_{ξ}$ -gauges}, JHEP 06
  (2017) 143.
\newblock \href {http://arxiv.org/abs/1704.03888} {\path{arXiv:1704.03888}},
  \href {https://doi.org/10.1007/JHEP06(2017)143}
  {\path{doi:10.1007/JHEP06(2017)143}}.

\bibitem{Alloul:2013bka}
A.~Alloul, N.~D. Christensen, C.~Degrande, C.~Duhr, B.~Fuks, {FeynRules 2.0 - A
  complete toolbox for tree-level phenomenology}, Comput. Phys. Commun. 185
  (2014) 2250--2300.
\newblock \href {http://arxiv.org/abs/1310.1921} {\path{arXiv:1310.1921}},
  \href {https://doi.org/10.1016/j.cpc.2014.04.012}
  {\path{doi:10.1016/j.cpc.2014.04.012}}.

\bibitem{Degrande:2011ua}
C.~Degrande, C.~Duhr, B.~Fuks, D.~Grellscheid, O.~Mattelaer, T.~Reiter, {UFO -
  The Universal FeynRules Output}, Comput. Phys. Commun. 183 (2012) 1201--1214.
\newblock \href {http://arxiv.org/abs/1108.2040} {\path{arXiv:1108.2040}},
  \href {https://doi.org/10.1016/j.cpc.2012.01.022}
  {\path{doi:10.1016/j.cpc.2012.01.022}}.

\bibitem{Alwall:2014hca}
J.~Alwall, R.~Frederix, S.~Frixione, V.~Hirschi, F.~Maltoni, O.~Mattelaer,
  H.~S. Shao, T.~Stelzer, P.~Torrielli, M.~Zaro, {The automated computation of
  tree-level and next-to-leading order differential cross sections, and their
  matching to parton shower simulations}, JHEP 07 (2014) 079.
\newblock \href {http://arxiv.org/abs/1405.0301} {\path{arXiv:1405.0301}},
  \href {https://doi.org/10.1007/JHEP07(2014)079}
  {\path{doi:10.1007/JHEP07(2014)079}}.

\bibitem{Gleisberg:2008ta}
T.~Gleisberg, S.~Hoeche, F.~Krauss, M.~Schonherr, S.~Schumann, F.~Siegert,
  J.~Winter, {Event generation with SHERPA 1.1}, JHEP 02 (2009) 007.
\newblock \href {http://arxiv.org/abs/0811.4622} {\path{arXiv:0811.4622}},
  \href {https://doi.org/10.1088/1126-6708/2009/02/007}
  {\path{doi:10.1088/1126-6708/2009/02/007}}.

\bibitem{Belyaev:2012qa}
A.~Belyaev, N.~D. Christensen, A.~Pukhov, {CalcHEP 3.4 for collider physics
  within and beyond the Standard Model}, Comput. Phys. Commun. 184 (2013)
  1729--1769.
\newblock \href {http://arxiv.org/abs/1207.6082} {\path{arXiv:1207.6082}},
  \href {https://doi.org/10.1016/j.cpc.2013.01.014}
  {\path{doi:10.1016/j.cpc.2013.01.014}}.

\bibitem{Kilian:2007gr}
W.~Kilian, T.~Ohl, J.~Reuter, {WHIZARD: Simulating Multi-Particle Processes at
  LHC and ILC}, Eur. Phys. J. C71 (2011) 1742.
\newblock \href {http://arxiv.org/abs/0708.4233} {\path{arXiv:0708.4233}},
  \href {https://doi.org/10.1140/epjc/s10052-011-1742-y}
  {\path{doi:10.1140/epjc/s10052-011-1742-y}}.

\bibitem{Christensen:2010wz}
N.~D. Christensen, C.~Duhr, B.~Fuks, J.~Reuter, C.~Speckner, {Introducing an
  interface between WHIZARD and FeynRules}, Eur. Phys. J. C72 (2012) 1990.
\newblock \href {http://arxiv.org/abs/1010.3251} {\path{arXiv:1010.3251}},
  \href {https://doi.org/10.1140/epjc/s10052-012-1990-5}
  {\path{doi:10.1140/epjc/s10052-012-1990-5}}.

\bibitem{Hahn:2000kx}
T.~Hahn, {Generating Feynman diagrams and amplitudes with FeynArts 3}, Comput.
  Phys. Commun. 140 (2001) 418--431.
\newblock \href {http://arxiv.org/abs/hep-ph/0012260}
  {\path{arXiv:hep-ph/0012260}}, \href
  {https://doi.org/10.1016/S0010-4655(01)00290-9}
  {\path{doi:10.1016/S0010-4655(01)00290-9}}.

\bibitem{Shtabovenko:2016sxi}
V.~Shtabovenko, R.~Mertig, F.~Orellana, {New Developments in FeynCalc 9.0},
  Comput. Phys. Commun. 207 (2016) 432--444.
\newblock \href {http://arxiv.org/abs/1601.01167} {\path{arXiv:1601.01167}},
  \href {https://doi.org/10.1016/j.cpc.2016.06.008}
  {\path{doi:10.1016/j.cpc.2016.06.008}}.

\bibitem{Hahn:2010zi}
T.~Hahn, {Feynman Diagram Calculations with FeynArts, FormCalc, and LoopTools},
  PoS ACAT2010 (2010) 078.
\newblock \href {http://arxiv.org/abs/1006.2231} {\path{arXiv:1006.2231}},
  \href {https://doi.org/10.22323/1.093.0078} {\path{doi:10.22323/1.093.0078}}.

\bibitem{Aebischer:2017ugx}
J.~Aebischer, et~al., {WCxf: an exchange format for Wilson coefficients beyond
  the Standard Model}, Comput. Phys. Commun. 232 (2018) 71--83.
\newblock \href {http://arxiv.org/abs/1712.05298} {\path{arXiv:1712.05298}},
  \href {https://doi.org/10.1016/j.cpc.2018.05.022}
  {\path{doi:10.1016/j.cpc.2018.05.022}}.

\bibitem{Dedes:2018seb}
A.~Dedes, M.~Paraskevas, J.~Rosiek, K.~Suxho, L.~Trifyllis, {The decay $h\to
  \gamma\gamma$ in the Standard-Model Effective Field Theory}, JHEP 08 (2018)
  103.
\newblock \href {http://arxiv.org/abs/1805.00302} {\path{arXiv:1805.00302}},
  \href {https://doi.org/10.1007/JHEP08(2018)103}
  {\path{doi:10.1007/JHEP08(2018)103}}.

\bibitem{Dedes:2019bew}
A.~Dedes, K.~Suxho, L.~Trifyllis, {The decay $h\to Z \gamma$ in the
  Standard-Model Effective Field Theory}\href {http://arxiv.org/abs/1903.12046}
  {\path{arXiv:1903.12046}}.

\bibitem{Dawson:2018pyl}
S.~Dawson, P.~P. Giardino, {Higgs decays to $ZZ$ and $Z\gamma$ in the standard
  model effective field theory: An NLO analysis}, Phys. Rev. D97~(9) (2018)
  093003.
\newblock \href {http://arxiv.org/abs/1801.01136} {\path{arXiv:1801.01136}},
  \href {https://doi.org/10.1103/PhysRevD.97.093003}
  {\path{doi:10.1103/PhysRevD.97.093003}}.

\bibitem{Vryonidou:2018eyv}
E.~Vryonidou, C.~Zhang, {Dimension-six electroweak top-loop effects in Higgs
  production and decay}, JHEP 08 (2018) 036.
\newblock \href {http://arxiv.org/abs/1804.09766} {\path{arXiv:1804.09766}},
  \href {https://doi.org/10.1007/JHEP08(2018)036}
  {\path{doi:10.1007/JHEP08(2018)036}}.

\bibitem{Hesari:2018ssq}
H.~Hesari, H.~Khanpour, M.~Mohammadi~Najafabadi, {Study of Higgs Effective
  Couplings at Electron-Proton Colliders}, Phys. Rev. D97~(9) (2018) 095041.
\newblock \href {http://arxiv.org/abs/1805.04697} {\path{arXiv:1805.04697}},
  \href {https://doi.org/10.1103/PhysRevD.97.095041}
  {\path{doi:10.1103/PhysRevD.97.095041}}.

\bibitem{Dawson:2018liq}
S.~Dawson, P.~P. Giardino, {Electroweak corrections to Higgs boson decays to
  $\gamma\gamma$ and $W^+W^-$ in standard model EFT}, Phys. Rev. D98~(9) (2018)
  095005.
\newblock \href {http://arxiv.org/abs/1807.11504} {\path{arXiv:1807.11504}},
  \href {https://doi.org/10.1103/PhysRevD.98.095005}
  {\path{doi:10.1103/PhysRevD.98.095005}}.

\bibitem{Dawson:2018jlg}
S.~Dawson, A.~Ismail, {Standard model EFT corrections to Z boson decays}, Phys.
  Rev. D98~(9) (2018) 093003.
\newblock \href {http://arxiv.org/abs/1808.05948} {\path{arXiv:1808.05948}},
  \href {https://doi.org/10.1103/PhysRevD.98.093003}
  {\path{doi:10.1103/PhysRevD.98.093003}}.

\bibitem{Baglio:2018bkm}
J.~Baglio, S.~Dawson, I.~M. Lewis, {NLO Effects in EFT Fits to $W^+W^-$
  Production at the LHC}, Phys. Rev. D99~(3) (2019) 035029.
\newblock \href {http://arxiv.org/abs/1812.00214} {\path{arXiv:1812.00214}},
  \href {https://doi.org/10.1103/PhysRevD.99.035029}
  {\path{doi:10.1103/PhysRevD.99.035029}}.

\bibitem{Dawson:2018dxp}
S.~Dawson, P.~P. Giardino, A.~Ismail, {SMEFT and the Drell-Yan Process at High
  Energy}\href {http://arxiv.org/abs/1811.12260} {\path{arXiv:1811.12260}}.

\bibitem{Silvestrini:2018dos}
L.~Silvestrini, M.~Valli, {Model-independent Bounds on the Standard Model
  Effective Theory from Flavour Physics}\href {http://arxiv.org/abs/1812.10913}
  {\path{arXiv:1812.10913}}.

\bibitem{Neumann:2019kvk}
T.~Neumann, Z.~E. Sullivan, {Off-shell single-top-quark production in the
  Standard Model Effective Field Theory}\href {http://arxiv.org/abs/1903.11023}
  {\path{arXiv:1903.11023}}.

\bibitem{Aebischer:2018bkb}
J.~Aebischer, J.~Kumar, D.~M. Straub, {Wilson: a Python package for the running
  and matching of Wilson coefficients above and below the electroweak scale},
  Eur. Phys. J. C78~(12) (2018) 1026.
\newblock \href {http://arxiv.org/abs/1804.05033} {\path{arXiv:1804.05033}},
  \href {https://doi.org/10.1140/epjc/s10052-018-6492-7}
  {\path{doi:10.1140/epjc/s10052-018-6492-7}}.

\bibitem{Celis:2017hod}
A.~Celis, J.~Fuentes-Martin, A.~Vicente, J.~Virto, {DsixTools: The Standard
  Model Effective Field Theory Toolkit}, Eur. Phys. J. C77~(6) (2017) 405.
\newblock \href {http://arxiv.org/abs/1704.04504} {\path{arXiv:1704.04504}},
  \href {https://doi.org/10.1140/epjc/s10052-017-4967-6}
  {\path{doi:10.1140/epjc/s10052-017-4967-6}}.

\bibitem{Criado:2017khh}
J.~C. Criado, {MatchingTools: a Python library for symbolic effective field
  theory calculations}, Comput. Phys. Commun. 227 (2018) 42--50.
\newblock \href {http://arxiv.org/abs/1710.06445} {\path{arXiv:1710.06445}},
  \href {https://doi.org/10.1016/j.cpc.2018.02.016}
  {\path{doi:10.1016/j.cpc.2018.02.016}}.

\bibitem{Brivio:2017btx}
I.~Brivio, Y.~Jiang, M.~Trott, {The SMEFTsim package, theory and tools}, JHEP
  12 (2017) 070.
\newblock \href {http://arxiv.org/abs/1709.06492} {\path{arXiv:1709.06492}}.

\bibitem{Bakshi:2018ics}
S.~Das~Bakshi, J.~Chakrabortty, S.~K. Patra, {CoDEx: Wilson coefficient
  calculator connecting SMEFT to UV theory}, Eur. Phys. J. C79~(1) (2019) 21.
\newblock \href {http://arxiv.org/abs/1808.04403} {\path{arXiv:1808.04403}},
  \href {https://doi.org/10.1140/epjc/s10052-018-6444-2}
  {\path{doi:10.1140/epjc/s10052-018-6444-2}}.

\bibitem{Gabelmann:2018axh}
M.~Gabelmann, M.~Mühlleitner, F.~Staub, {Automatised matching between two
  scalar sectors at the one-loop level}, Eur. Phys. J. C79~(2) (2019) 163.
\newblock \href {http://arxiv.org/abs/1810.12326} {\path{arXiv:1810.12326}},
  \href {https://doi.org/10.1140/epjc/s10052-019-6570-5}
  {\path{doi:10.1140/epjc/s10052-019-6570-5}}.

\bibitem{Bishara:2017nnn}
F.~Bishara, J.~Brod, B.~Grinstein, J.~Zupan, {DirectDM: a tool for dark matter
  direct detection}\href {http://arxiv.org/abs/1708.02678}
  {\path{arXiv:1708.02678}}.

\bibitem{Denner:1992vza}
A.~Denner, H.~Eck, O.~Hahn, J.~Kublbeck, {Feynman rules for fermion number
  violating interactions}, Nucl. Phys. B387 (1992) 467--481.
\newblock \href {https://doi.org/10.1016/0550-3213(92)90169-C}
  {\path{doi:10.1016/0550-3213(92)90169-C}}.

\bibitem{Denner:1992me}
A.~Denner, H.~Eck, O.~Hahn, J.~Kublbeck, {Compact Feynman rules for Majorana
  fermions}, Phys. Lett. B291 (1992) 278--280.
\newblock \href {https://doi.org/10.1016/0370-2693(92)91045-B}
  {\path{doi:10.1016/0370-2693(92)91045-B}}.

\bibitem{Paraskevas:2018mks}
M.~Paraskevas, {Dirac and Majorana Feynman Rules with four-fermions}\href
  {http://arxiv.org/abs/1802.02657} {\path{arXiv:1802.02657}}.

\bibitem{Jenkins:2013zja}
E.~E. Jenkins, A.~V. Manohar, M.~Trott, {Renormalization Group Evolution of the
  Standard Model Dimension Six Operators I: Formalism and lambda Dependence},
  JHEP 10 (2013) 087.
\newblock \href {http://arxiv.org/abs/1308.2627} {\path{arXiv:1308.2627}},
  \href {https://doi.org/10.1007/JHEP10(2013)087}
  {\path{doi:10.1007/JHEP10(2013)087}}.

\bibitem{Jenkins:2013wua}
E.~E. Jenkins, A.~V. Manohar, M.~Trott, {Renormalization Group Evolution of the
  Standard Model Dimension Six Operators II: Yukawa Dependence}, JHEP 01 (2014)
  035.
\newblock \href {http://arxiv.org/abs/1310.4838} {\path{arXiv:1310.4838}},
  \href {https://doi.org/10.1007/JHEP01(2014)035}
  {\path{doi:10.1007/JHEP01(2014)035}}.

\bibitem{Alonso:2013hga}
R.~Alonso, E.~E. Jenkins, A.~V. Manohar, M.~Trott, {Renormalization Group
  Evolution of the Standard Model Dimension Six Operators III: Gauge Coupling
  Dependence and Phenomenology}, JHEP 04 (2014) 159.
\newblock \href {http://arxiv.org/abs/1312.2014} {\path{arXiv:1312.2014}},
  \href {https://doi.org/10.1007/JHEP04(2014)159}
  {\path{doi:10.1007/JHEP04(2014)159}}.

\bibitem{Vermaseren:1994je}
J.~A.~M. Vermaseren, {Axodraw}, Comput. Phys. Commun. 83 (1994) 45--58.
\newblock \href {https://doi.org/10.1016/0010-4655(94)90034-5}
  {\path{doi:10.1016/0010-4655(94)90034-5}}.

\end{thebibliography}


\begin{thebibliography}{0}
\bibitem{1} A.~Dedes, W.~Materkowska, M.~Paraskevas, J.~Rosiek and
  K.~Suxho, ``Feynman rules for the Standard Model Effective Field
  Theory in R$_{\xi}$-gauges,'' JHEP {\bf 1706}, 143 (2017),
  arXiv:1704.03888.
\bibitem{2} A.~Alloul, N.~D.~Christensen, C.~Degrande, C.~Duhr and
  B.~Fuks, ``FeynRules 2.0 - A complete toolbox for tree-level
  phenomenology,'' Comput.\ Phys.\ Commun.\ {\bf 185}, 2250 (2014),
  arXiv:1310.1921.

  
\end{thebibliography}

\end{document}